\documentclass[a4paper]{jpconf}
\usepackage{graphicx}
\usepackage{amsmath,amssymb}
\usepackage{url}

\begin{document}
	\title{Aspects of Axions and ALPs Phenomenology}
	
	\author{Maurizio Giannotti}
	
	\address{Department of Chemistry and Physics, Barry University, 11300 NE 2nd Ave., Miami Shores, FL 33161, USA}
	
	\ead{mgiannotti@barry.edu}
	
	\begin{abstract}
		The physics of axions and axion-like particles (ALPs) is enjoying an incredibly productive period, with many new experimental proposals, 
		theoretical idea, and original astrophysical and cosmological arguments which help the search effort. 
		The large number of experimental proposals is likely to lead to fundamental advances (perhaps, a discovery?) in the coming years. 
		The aim of this article is to provide a very brief overview of some of the recent developments in axions and ALP phenomenology, 
		and to discuss some relevant aspects in this important field.
		A particular attention is given to the definition of  
		motivated regions in the axion parameters space, which should be the targets of experimental searches. 
	\end{abstract}
	
	\section{Introduction}
	Axions and Axion Like Particles (ALPs) are ubiquitous in modern theoretical physics, as they emerge in several extensions of the Standard Model of particle physics (SM)~\cite{Ringwald:2014vqa}.
	Generally, the term ALP applies to  pseudoscalar particles described by the low energy phenomenological Lagrangian 
	\begin{eqnarray}
		{\mathcal L}_{a}=\frac12 (\partial_\mu a)^2 
		-m_a^2 a^2  - \sum_{f=e,p,n} g_{af} a \bar{\psi}_f\gamma_5\psi_f
		-\frac14 g_{a\gamma} \, a\,F_{\mu\nu} \tilde{F}^{\mu\nu}\,,
		\label{eq:L_a}
	\end{eqnarray}
	where $a$ is the ALP field with mass $m_a$, $\psi_f$  represent the electron and nucleon fields, 
	$F$ is the electromagnetic field strength tensor and $\tilde F$ its dual.
	In such a description, the ALP interactions with the SM are parameterized by a set of  coupling constants $g_{ai}$ $(i=f, \gamma)$.\footnote{We ignore interactions with fermions other than $e$, $p$ and $n$, as they are less relevant for axion searches. We also neglect possible CP-odd couplings~\cite{OHare:2020wah}.}
	
	Arguably, the most well motivated ALP model is the QCD axion~\cite{Weinberg:1977ma,Wilczek:1977pj}, often known simply as axion, a prediction of the Peccei and Quinn (PQ) solution of the strong CP problem~\cite{Peccei:1977hh,Peccei:1977ur}. 
	Besides their role in particle physics, axions are also favorite dark matter (DM) candidates (see Refs.~\cite{Marsh:2015xka,DiLuzio:2020wdo,Adams:2022pbo} for recent reviews). 
	We shall review some general features of the QCD axion in Sec.~\ref{sec:QCD_axion}.
	
	Besides QCD axions, light pseudoscalar particles emerge also in other scenarios which extend the SM. 
	For example, they are a common prediction of string theory~\cite{Svrcek:2006yi,Arvanitaki:2009fg,Cicoli:2012sz}. 
	These ALPs share a lot of the characteristics of the QCD axion but are, in general, much less constrained. For example, they do not require specific relations between mass and couplings. 
	
	The experimental search for axions and ALPs has enjoyed a fantastic revival in the last few years (see Refs.~\cite{Irastorza:2018dyq,Sikivie:2020zpn,Irastorza:2021tdu} for recent reviews). 
	In particular, new experimental techniques promise the exploration of large, currently unconstrained portions of the ALP parameter space. 
	A groundbreaking discovery in the coming decade or so is not such a remote possibility. 
	Furthermore, as we shall discuss below, large portions of the region accessible to modern experiments are well motivated by astrophysical and cosmological considerations. 
	The significant role of ALPs in particle physics, astrophysics and cosmology contributes to the current enthusiasm for this research field~\cite{Giannotti:2017law}.  
	Here, I provide a very brief review of some key aspects of axions and ALP phenomenology. 
	
	\section{A few facts about the QCD axion}
	\label{sec:QCD_axion}
	Historically, QCD axions emerged from an attempt to explain the observed time-reversal (hence, CP) symmetry of the strong interactions.
	At the origin of this problem, known as the strong CP problem, is the CP-odd $\theta$-term in the QCD Lagrangian 
	$L_{\theta} =(\theta/32\pi^2) G\tilde G$, with $G$ the gluon field.
	The parameter $\theta$ can be measured, for example, through its contribution to the neutron electric dipole moment~\cite{Baluni:1978rf,Crewther:1979pi,Pospelov:1999mv}, 
	which has been measured with higher and higher precision since the 1950s~\cite{Smith:1957ht}. 
	Present measurements~\cite{Abel:2020gbr} require $\theta <10^{-10}$ or so, an extremely low value for a parameter which has no fundamental reason to be zero.\footnote{The contribution to $\theta$ from the weak sector is negligible, of order $ 10^{-17}$~\cite{Georgi:1986kr,Ellis:1978hq}.}
	
	From a modern point of view, the PQ mechanism solves the strong CP problem by promoting $ \theta $ from a constant parameter to a field, the axion $a$:
	\begin{equation}
		\label{eq:}
		\frac{g_s^2}{32\pi^2}\, \theta \,G^a_{\mu\nu}\tilde G^{a\mu\nu} 
		\to
		\frac12 (\partial_\mu a)^2 + \frac{g_s^2}{32\pi^2}
		\left( \theta+\frac{a}{f_a} \right) 
		\, G^a_{\mu\nu}\tilde G^{a\mu\nu} \,,
	\end{equation}
	where $f_a$ is a model dependent energy scale known as the PQ or axion constant.
	It is possible to show that the energy of vector-like gauge theories such as QCD is minimized at the CP conserving point~\cite{Vafa:1984xg}. 
	In this sense, the PQ mechanism offers a \textit{dynamical solution} of the strong CP problem: 
	since the axion is a dynamical degree of freedom, it will settle in the CP conserving minimum of the potential.
	
	Notice that the introduction of the new energy constant, $f_a$, is required by simple dimensional analysis, to compensate for the dimension of the new scalar field.
	Therefore, the original $\theta$-term is transformed into a dimension 5 operator, implying that the theory needs a UV completion.
	The common realization of the PQ mechanism proceed through the introduction of a new abelian axial quasi-symmetry (the PQ symmetry), broken by the color anomaly and spontaneously broken at the PQ scale $f_a$. 
	The axion is therefore interpreted as the (pseudo)Goldstone boson of this symmetry. 
	Depending on the UV completion, there are several possible theories, with very specific features and often very different phenomenological predictions. 
	A comprehensive review can be found in Ref.~\cite{DiLuzio:2020wdo}.
	Nevertheless, there are some features that are more or less general and apply to most QCD axion models:
	\begin{enumerate}
		\item The axion potential can be calculated in QCD~\cite{GrillidiCortona:2015jxo,Borsanyi:2016ksw}. 
		In particular, QCD fixes the axion mass for any given value of $f_a$~\cite{GrillidiCortona:2015jxo}
		\begin{equation}
			\label{eq:ma_fa}
			m_a=\frac{(5.7\pm 0.07)\times 10^6}{f_a/{\rm GeV}}\,{\rm eV} \,.
		\end{equation}
		This relation can be relaxed in specific axion models but doing so requires significant changes to the standard QCD, for example the addition of a new QCD sector~\cite{Rubakov:1997vp,Berezhiani:2000gh,Gianfagna:2004je}. 
		Several possibilities are discussed in Ref.~\cite{DiLuzio:2020wdo}.
		\item Symmetry considerations dictate the form of the axion interaction Lagrangian, as given in Eq.~(\ref{eq:L_a}).
		The coupling constants, however, cannot be fully predicted because of the substantial corrections from the model dependent UV contribution~\cite{GrillidiCortona:2015jxo}. 
		It is in fact possible, at least in principle, to conceive axion models with very small couplings to photons and matter fields, and therefore very difficult to detect (see, e. g., Refs.~\cite{DiLuzio:2017ogq,Bjorkeroth:2019jtx}, and appendix A of Ref.~\cite{Lucente:2022vuo}).
		\item Though it is in principle possible to conceive models that are extremely weakly interacting, 
		axions cannot be completely invisible. 
		\textit{First}, they must interact with gravity in a universal way. 
		Thus, for example, the non-thermal contributions to the axion DM density are largely model independent. 
		The same applies to the astrophysical constraints from black hole superradiance~\cite{Arvanitaki:2010sy,Arvanitaki:2014wva,Cardoso:2018tly}. 
		\textit{Second}, the gluonic vertex, required for the solution of the strong CP problem, induces a model independent \emph{nucleon EDM portal} interaction
		\begin{equation}
			L_a^{\rm nEDM} =-\frac{i}{2} g_{d,N} a {\bar N} \gamma_5 \sigma_{\mu\nu} N {F}^{\mu\nu} \, ,
			\label{eq:dipole portal}
		\end{equation}
		(with $N = p,n$ and $g_{d,n}=-g_{d,p}\equiv g_{d}$) which leads to an axion-dependent nucleon electric dipole moment (EDM), $d_N=g_{d,N} a$~\cite{Graham:2013gfa}. 
		\item Axions contribute at least a fraction of the DM of the universe~\cite{Adams:2022pbo}. 
		Unfortunately, however, it is very difficult to estimate the exact fraction of axion DM. 
	\end{enumerate}

	\begin{figure}[t]
		\begin{minipage}{18pc}
			\includegraphics[width=18pc]{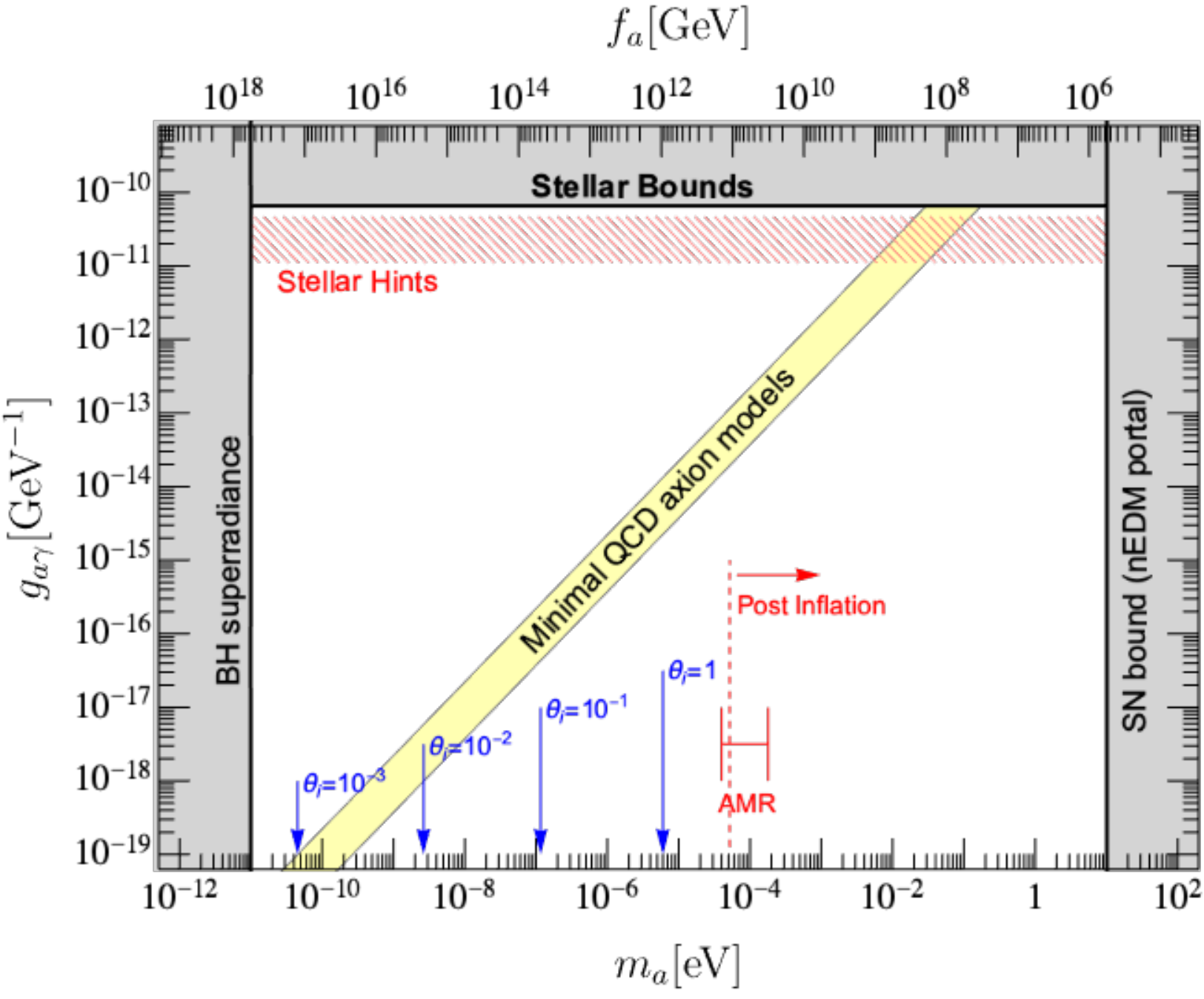}
			\caption{\label{fig_axion_space} Axion parameter space allowed under the solely assumption that axion mass and constant are related as in Eq.~\eqref{eq:ma_fa}.
				The yellow band indicates the range of parameters for a set of minimal assumptions (see text for more details). 
				The arrows indicate the value of the mass required to guarantee that axions contribute to the totality of the DM in the universe (data from Ref.~\cite{Borsanyi:2016ksw}).
				The red lines refer to the post-inflationary scenario. The mass limit is from Ref.~\cite{Borsanyi:2016ksw} while the hinted mass region, labeled AMR, is from Ref.~\cite{Buschmann:2021sdq}.}
		\end{minipage}\hspace{2pc}%
		\begin{minipage}{18pc}
			\includegraphics[width=18pc]{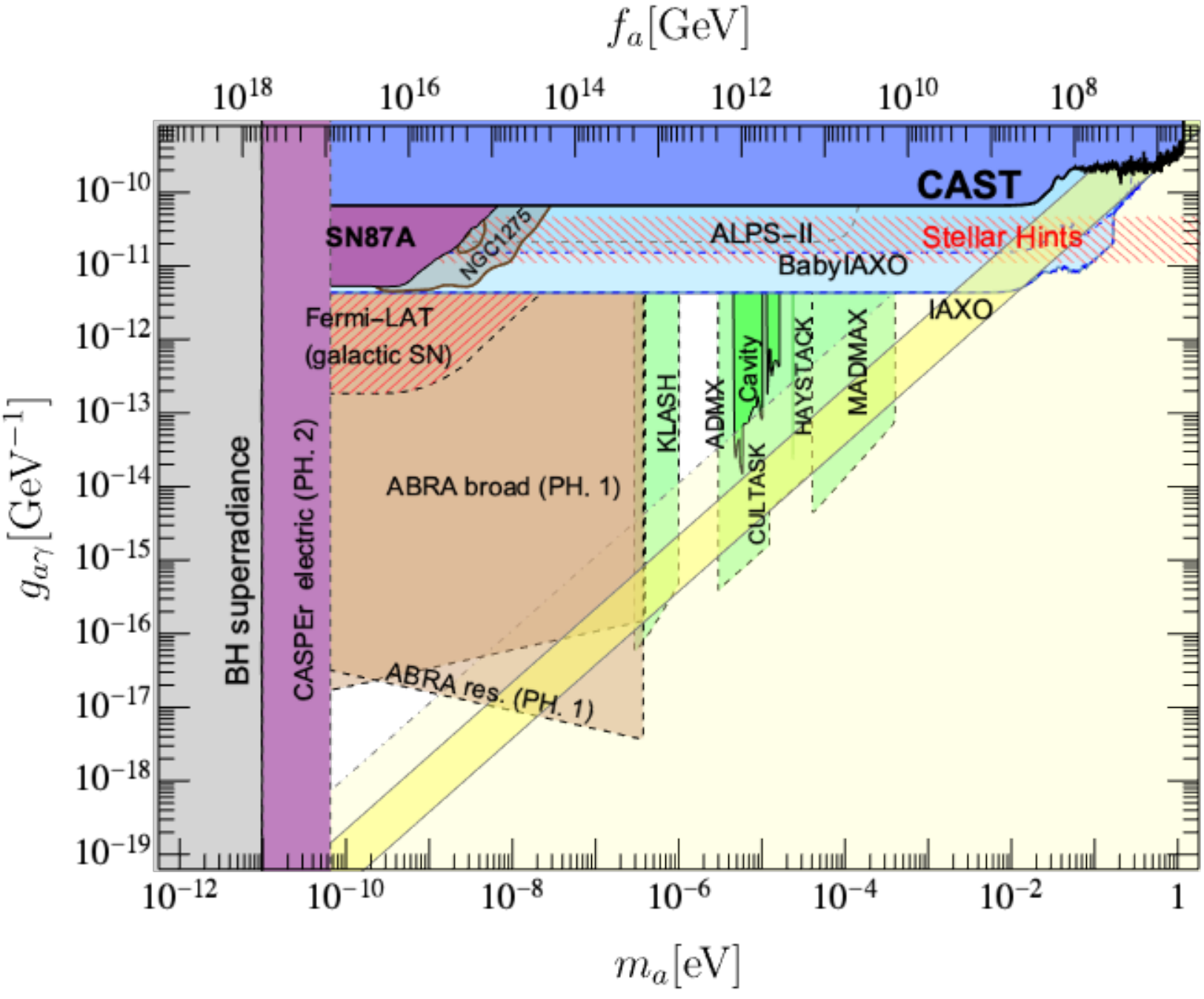}
			\caption{\label{fig_axion_space_complete} Same as figure~\ref{fig_axion_space}, but including current and future axion searches (adapted from Ref.~\cite{DiLuzio:2020wdo}).
				Current experimental bounds and phenomenological constraints are included in continuous lines. 
				Sensitivity of future searches are shown in dashed lines. 
				The darker yellow line defines the minimal QCD axion models, as in figure~\ref{fig_axion_space}. 
				The lighter yellow indicates the parameter space for hadronic axions with a non-minimal choice of the fermion representation, following Ref.~\cite{DiLuzio:2016sbl,DiLuzio:2017pfr}.  
			}
		\end{minipage} 
	\end{figure}
	
	\section{Where are the axions?}
	Some fairly general considerations allow to define a mass region preferred by the simplest axion models. 
	
	To start, we may assume that the relation between mass and PQ constant in Eq.~(\ref{eq:ma_fa}) is satisfied, 
	since relaxing it would require non-minimal assumptions (see Sec.~\ref{sec:beyond_QCD_band}).
	In this case, asking for the PQ scale to be below the Planck scale implies $m_a<$ a few $10^{-13}\,$eV.
	A stronger constraint derives from the observation of black hole spins at fixed masses. 
	Because of black hole superradiance, very light axions could subtract spin to black holes, spoiling the observations~\cite{Arvanitaki:2010sy,Arvanitaki:2014wva,Cardoso:2018tly}. 
	This argument can be used to constrain $m_a\gtrsim 10^{-11}\,$eV~\cite{Cardoso:2018tly}.
	
	Analogously, a model independent upper bound on the axion mass is derived from the effects of the axion-nEDM coupling in the cooling of supernovae~\cite{Graham:2013gfa}.
	The most recent analysis provides the bound $m_a\lesssim 10.1\,$eV~\cite{Lucente:2022vuo}.\footnote{Masses larger than $\sim 12$ keV are also allowed by the supernova argument, since such axions would be trapped in the SN core~\cite{Lucente:2022vuo}. However, these masses require a quite large PQ constant and it is quite more difficult to conceive models which are sufficiently weakly coupled to avoid observational constraints. 
		We ignore this case in the name of simplicity, though in principle the possibility is phenomenologically viable. }
	
	Let's now focus on the axion-photon coupling, $g_{a\gamma}$, which is arguably the most studied axion coupling. 
	Astrophysical considerations show that $g_{a\gamma}<6.5\times 10^{-11}\,$GeV in the entire mass range presented above~\cite{Ayala:2014pea,Straniero:2015nvc}.
	The parameter space derived by the above simple arguments is shown in figure~\ref{fig_axion_space}.
	It spans 12 orders of magnitude in mass and, as discussed later, it may also be extended beyond these boundaries in non-minimal axion models. 
	Some further (and less general) considerations can help reduce this space. 
	
	Theoretical arguments about the \textit{simplicity} of the models allow to select an \textit{axion window} in this very large box.  
	In general, even the simplest axion model requires the addition of fields beyond the standard model (SM). 
	There are two minimal options: 
	\textit{1)} enlarge the SM with a set of heavy colored fermions and a singlet scalar field or 
	\textit{2)} enlarge the SM with an additional Higgs field and a singlet scalar field.
	The first option defines the so called hadronic axion models, whose benchmark example is the KSVZ axion~\cite{Kim:1979if,Shifman:1979if}, while the most significant representative of the second category is the DFSZ axion~\cite{Zhitnitsky:1980tq,Dine:1981rt}.
	A set of theoretical and phenomenological considerations allow to restrict the range of photon couplings associated with each mass.
	In particular, defining $g_{a\gamma}=\alpha\,C_{a\gamma}/2\pi f_a$, it is possible to show a phenomenologically preferred axion window for $0.25\leq C_{a\gamma}\leq 12.7$~\cite{DiLuzio:2017pfr,DiLuzio:2017pfr}.
	In the case of non-hadronic models, the requirement of only one additional Higgs field implies  $0.08\leq C_{a\gamma}\leq 5.25$~\cite{DiLuzio:2021ysg}.
	These requirements select a band in the $ g_{a\gamma}-m_a $ plane, usually referred to as the \textit{QCD axion window}, shown in yellow in figure~\ref{fig_axion_space}.
	It is, quite possible to design axion models that live outside the axion window.
	However, in general going below that window requires some tuning between independent contributions to the axion-photon coupling~\cite{Kaplan:1985dv} 
	while increasing the coupling above the preferred window may require the addition of several other fields (see, e.g., Ref.~\cite{Darme:2020gyx}).
	
	Even if we take the above simplicity argument as a guide, the axion parameter space remains quite large. 
	We have to turn to phenomenology to help defining target regions that are particularly well motivated.

	\subsection{What can we learn from cosmology} 
	\label{sec:cosmology}
	The topic of axion (and ALP) cosmology is extremely interesting and rich.
	Among the many excellent reviews, one of the most comprehensive is Ref.~\cite{Marsh:2015xka} while the most updated results are summarized in the reports~\cite{Adams:2022pbo,Berti:2022rwn}.
	In the spirit of this paper, here we avoid details and technical discussions, pointing out instead to some general features, problems, and results of interest for future searches. 
	
	The first observation is that for the PQ mechanism to work, axion must be at least a fraction of the cold DM (CDM) in the universe. 
	What fraction, exactly, depends on the axion mass (not on its couplings). 
	In the spirit of elegance and simplicity, it would be obviously very appealing if axions were, indeed, the totality of the CDM in the universe. 
	Let's refer to this as the ``\textit{axion CDM condition}".
	The axion parameter space where this condition is satisfied should be an extremely well motivated target for experimental searches. 
	Indeed, historically this has guided (and keeps guiding) important experimental decisions. 
	Calculating (or even estimating) the axion mass corresponding to the CDM condition is, however, an outstanding problem, still far from being resolved. 
	
	But let's proceed in steps. 
	As we mentioned above, the PQ mechanism does not solve the strong CP problem instantaneously. 
	After the $U(1)_{PQ}$ is spontaneously broken, at the energy scale $f_a$, the $\theta$ angle has a random initial value, $\theta_i$.
	It is only later, when the axion acquires a mass comparable to the Hubble friction, that the axion field begins to oscillate around the CP-conserving minimum of the potential, 
	generating extremely low momentum axions, which behave as CDM.
	These oscillations are an inevitable consequence of the PQ mechanism. 
	If the axion mass is such that they never start, the strong CP would not be solved.
	If they do start, they are still going on, since the universe has a finite age. 
	This is known as the misalignment (or vacuum realignment) mechanism and it was first introduced in Refs.~\cite{Abbott:1982af,Dine:1982ah,Preskill:1982cy}. 
	A recent analysis is provided in Ref.~\cite{Borsanyi:2016ksw}, which is the one we refer to here, for numerical results. 
	The axion energy density associated with this process depends not only on the axion mass but also on the (in general unknown) value of the initial $ \theta $
	angle, $\theta_i$
	\begin{align} 
		\Omega_{a} h^2 
		\sim
		0.1
		\left(
		\frac{f_a} {10^{12}~{\rm GeV}} 
		\right)^{7/6} \theta_i^2 
		\label{eq:axion-abundance}
		\,.
	\end{align} 
	The value of $\theta_i$ depends strongly on the cosmological history. 
	
	In the \textit{pre-inflationary} scenario, the PQ mechanism happens before inflation and the different causally disconnected patches with a random value of $ \theta_i $ get largely inflated.
	In this case, our universe today is generated from a patch with a single random value of $\theta_i$, which fixes
	the mass satisfying the axion CDM condition. 
	This is shown with the blue arrows in figure~\ref{fig_axion_space}.
	
	In the \textit{post-inflationary} scenario, on the other hand, the PQ mechanism follows inflation and so the present universe encloses a large number of patches with different random values of $ \theta_i $.
	In this case, the problem of the unknown initial $ \theta $ value disappears since 
	the axion abundance depends on the average value of the $ \theta_i $ in the different patches.
	However, even in this case the axion mass corresponding to the axion CDM condition cannot be extracted with confidence, because of other 
	effective axion production mechanisms available in the post-inflationary scenario. 
	It is known, in fact, that phase transitions generate topological defects, in particular axion strings~\cite{Kibble:1976sj,Kibble:1982dd} and these, in turn, may produce axions~\cite{Davis:1986xc,Harari:1987ht}.
	These topological defects do not play a role in the pre-inflationary scenario (they would be inflated away), 
	but may be very relevant if inflation precedes the PQ mechanism. 
	The calculation of the axion abundance from topological defects is extremely challenging, numerically, mainly because of the many orders of magnitude difference in the size of the strings core and their mutual distances. 
	A significant progress was achieved recently through the use of the adaptive mesh refinement (AMR)
	technique~\cite{Drew:2019mzc}, which led to the prediction of the mass window $m_a\in [40,180]\,\mu$eV for the axion to be the totality of the CDM in the universe~\cite{Buschmann:2021sdq}.
	Nevertheless, more work is required for a complete confidence in these results. 
	It is also worth noting that previous simulations hinted at quite different mass region (see, e.g., Refs.~\cite{Gorghetto:2018myk,Gorghetto:2020qws}).
	In any case, the recent progress provides ground for optimism in the selection of a cosmologically preferred mass region. 
	A comprehensive and updated review, including a revised bibliography, can be found in Ref.~\cite{OHare:2021zrq}.

	\subsection{What can we learn from stellar evolution} 
	Just like other light and weakly interacting particles, also known as FIPs (Feebly Interacting Particles)~\cite{Agrawal:2021dbo}, axions and ALPs can be produced in stars and, for a wide range of the parameters, free stream out of them.
	This makes stars extremely efficient axion (and ALP) factories. 
	The Sun is the most studied example and one of the most important axion (and ALPs) sources in nature. 
	A few recent studies and results concerning the axion coupling with photons, electrons and nucleons can be found in Refs.~\cite{Redondo:2013wwa,Hoof:2021mld,CAST:2017uph,Avignone:2017ylv,Bhusal:2020bvx,DiLuzio:2021qct}. 
	
	Besides providing a source for axions, stars are also very efficient laboratories to study their properties~\cite{Raffelt:1996wa}. 
	Specifically, stellar observations allow to set stringent bounds on the axion couplings with the standard model fields (an updated summary can be found in Ref.~\cite{DiLuzio:2021ysg}).
	The exclusion region labeled ``Stellar Bounds" in figure~\ref{fig_axion_space} encloses the axion parameters which would accelerate the evolution of horizontal branch (HB) stars, and thus reduce their population in globular clusters, beyond the observational constraints~\cite{Ayala:2014pea,Straniero:2015nvc}.
	Many other stars have been used to constrain the other axion couplings and, perhaps more intriguingly, several stellar systems indicate a preference for additional cooling that could be provided by axions or ALPs~\cite{Giannotti:2015kwo,Giannotti:2017hny,DiVecchia:2019ejf,DiLuzio:2021ysg}. 
	A summary of the stellar bounds and hints on the axion properties is provided in table~\ref{tab:Astro_bounds}.
	The hatched region in figure~\ref{fig_axion_space}, labeled ``Stellar Hints", refers to the HB-hint on the axion-photon coupling shown in the table. 
	\begin{table}[t]
		\caption{\label{tab:Astro_bounds}
			Bounds and hints from stellar evolution on the axion couplings (updated from Ref.~\cite{DiLuzio:2021ysg}). 
			Legend: $ g_{e13}\equiv g_{ae}/10^{-13} $; $ g_{\gamma 10}\equiv g_{a\gamma}/10^{-10} {\rm GeV}^{-1}$;
			WDLF: White Dwarf Luminosity Function; 
			WDV: White Dwarf Variables; 
			RGBT: Red Giant Branch Tip;
			HB: Horizontal Branch;
			SN: Supernova; NS: Neutron Star.
			In the case of SN, the more recent results in Refs.~\cite{Carenza:2020cis,Fischer:2021jfm} have not been translated into bound on the axion couplings yet. 
		}
		\begin{center}
			\begin{tabular}{ l l l l}
				\br
				Star  				&  Hint 										&  Bound  					& Reference	 		\\ \mr
				\ Sun \phantom{$\Big|$} &   No Hint 												    &  $g_{\gamma10}\leq 2.7$ 	&	\cite{Vinyoles:2015aba}	\\
				\vspace{0.2cm}
				WDLF 				&  $g_{e13}=1.5^{+0.3}_{-0.5}$ 					&  $g_{e13}\leq 2.1$ 	&	\cite{Bertolami:2014wua}  	\\
				\vspace{0.2cm}
				WDV 		 	 	&  $g_{e13}=2.9^{+0.6}_{-0.9}$ 		 			&  $g_{e13}\leq 4.1$ 	 &	 \cite{Corsico:2019nmr}	 \\
				\vspace{0.2cm}
				RGBT	(22 GGCs)	&  $g_{e13}=0.60^{+0.32}_{-0.58}$ 	            &  $g_{e13}\leq 1.5$        &     \cite{Straniero:2020iyi}    \\	 
				\vspace{0.2cm}
				RGBT	(NGC 4258)  &  No Hint 		         				                        &  $g_{e13}\leq 1.6$ 	&	\cite{Capozzi:2020cbu}  \\
				\vspace{0.2cm}
				HB 					&  $g_{\gamma10}=0.3^{+0.2}_{-0.2}$ 	&  $g_{\gamma10}\leq 0.65$ 	&	\cite{Ayala:2014pea,Straniero:2015nvc}\\
				\vspace{0.2cm}
				SN 1987A 			&  No Hint 	&  $ g_{an}^2+ 0.6 g_{ap}^2 + 0.5 g_{an}g_{ap}\lesssim 8.26 \times \! 10^{-19}   $ 		&	 \cite{Carenza:2019pxu}\\ 
				\vspace{0.2cm}
				NS (CAS A) 			&  No Hint 	&  $ (g_{ap}^2+1.6\, g_{an}^2)^{1/2}\lesssim 1.0\!\times\! 10^{-9} $ 		&	\cite{Hamaguchi:2018oqw} \\ 
				\vspace{0.2cm}
				NS (CAS A) 			&  No Hint 	&  $g_{an}\lesssim 3\!\times\! 10^{-10}$		&	\cite{Leinson:2021ety} \\ 
				\vspace{0.2cm}
				NS (HESS) 			&  No Hint 	&  $g_{an}\leq 2.8\!\times\! 10^{-10}$		&	\cite{Beznogov:2018fda} \\ 
				\vspace{0.2cm}
				NS (Magnificent 7) 		&  No Hint 	&  $g_{an},g_{ap}\lesssim 1.5\!\times\! 10^{-9}$		&	\cite{Buschmann:2021juv} \\ 
				\br
			\end{tabular} 
		\end{center}
	\end{table}

	A comprehensive global analysis of all the stellar bounds and hints showed that the axions couplings preferred by stars are $g_{a\gamma}\simeq 2\times 10^{-11}\,{\rm GeV}^{-1}$ and $g_{ae}\simeq 10^{-13}$~\cite{DiLuzio:2021ysg}.
	For axions in the QCD band, this corresponds to axion masses from a few meV to $ \sim $ 100 meV, in a region accessible mostly to 
	axion helioscopes~\cite{IAXO:2019mpb,IAXO:2020wwp,Galan:2015msa,Galan:2015caz}, but possibly also to new haloscope concepts such as 
	TOORAD~\cite{Marsh:2018dlj,Schutte-Engel:2021bqm} or BREAD~\cite{BREAD:2021tpx} (see also section~\ref{sec:future}).

	\subsection{Breaking the boundaries} 
	\label{sec:beyond_QCD_band}
	
	To be clear, axions are not required to live in the yellow QCD band of figure~\ref{fig_axion_space}. 
	The band is derived under the assumption of simplicity, not consistency. 
	Nevertheless, it is good to set guides based on simplicity and, in my opinion, it is wise to give more relevance to the experimental exploration of the parameter space corresponding to minimal models. 
	
	Said this, various possibilities of expanding the allowed QCD axion window have been explored for decades.\footnote{It is worth mentioning that recent investigations about the role of magnetically charged fermions in the axion dynamics~\cite{Sokolov:2021ydn,Sokolov:2022fvs} presented axion models coupled much more strongly to photons than the models in the QCD band. 
		We will not discuss this possibility here, and remind the interested reader to the original studies.} 
	Already in 1985, David B. Kaplan showed that axions may be very weakly coupled to photons, though at the price of some tuning. 
	More recently, in Ref.~\cite{DiLuzio:2016sbl,DiLuzio:2017pfr}, which, perhaps for the first time, presented a well motivated definition of the axion window,
	it was shown that hadronic models with heavy quarks in multiple representations of the SM are allowed in a considerably wider window. 
	This is shown in figure~\ref{fig_axion_space_complete}, shaded in light yellow. 
	
	A more aggressive approach can lead to quite significant departures from the axion window. 
	Mechanisms inspired by \textit{clockwork-type} constructions can provide exponentially enhanced axion couplings at fixed masses. 
	We are not entering into theoretical details here, and refer the reader to reviews such as Refs.~\cite{DiLuzio:2020wdo,Agrawal:2021dbo}. 
	We just mention that these mechanisms ask for several additional fields to the SM but, in general, do not require relaxing relation \eqref{eq:ma_fa}. 
	
	A conceptually different approach consists in relaxing the relation between mass and PQ constant, in Eq.~\eqref{eq:ma_fa}. 
	As mentioned above, this requires some modifications of the gauge structure of the SM,
	particularly of the strong interaction sector. 
	Such modifications demand, of course, some care in order not to
	shift the minimum of the axion potential away from its CP conserving point, thus spoiling 
	the solution of the strong CP problem. 
	Again, this is not the place to discuss the numerous options and ideas that were put forward in the past, starting from the earliest attempts in Refs~\cite{Holdom:1982ex,Holdom:1985vx}, to more recent ones in, e.g., Refs.~\cite{Hook:2014cda,Dimopoulos:2016lvn,Hook:2019qoh}.
	A comprehensive review, with a more complete bibliography, is presented in section 6.7 of Ref.~\cite{DiLuzio:2020wdo}.
	Here, we just want to mention that relaxing relation~\eqref{eq:ma_fa} would allow to conceive 
	\textit{superlight} as well as \textit{superheavy} axion models. 
	A significant improvement in building axion models much lighter than what predicted by Eq.~\eqref{eq:ma_fa} resulted from the mechanism proposed in Ref.~\cite{Hook:2018jle}. 
	This was exploited in Refs.~\cite{DiLuzio:2021pxd,DiLuzio:2021gos} to conceive models with exponentially suppressed mass, 
	which still solved the strong CP problem and saturated the dark matter abundance in the universe. 
	Superheavy axions would also have a rich phenomenology, although they would not likely be DM candidates (the lifetime scales as $m_a^{-3}$). 
	In particular, such models would not be confined by the 
	supernova nEDM bound, $ m_a\lesssim $ 10 eV, 
	and could expand beyond the wall on the right side of figure~\ref{fig_axion_space}.
	At masses larger than a few 100 keV, a lot of the astrophysical bounds, for example the one from HB stars shown in figure~\ref{fig_axion_space}, 
	would also rapidly weaken~\cite{Lucente:2022wai}, opening up the space for experimental 
	searches.\footnote{Some bounds from SN 1987A would still remain in place, limiting the regions accessible to experiments. See, e.g., Ref.~\cite{Caputo:2021rux} for a recent analysis.} 
	Several experiments are already probing that region of the parameter space (see, e.g., \cite{Agrawal:2021dbo} and references therein).

	\section{A bird look at the ALP parameter space}
	\label{sec:parameter_space}
	
	\begin{figure}[t]
		\includegraphics[width=22pc]{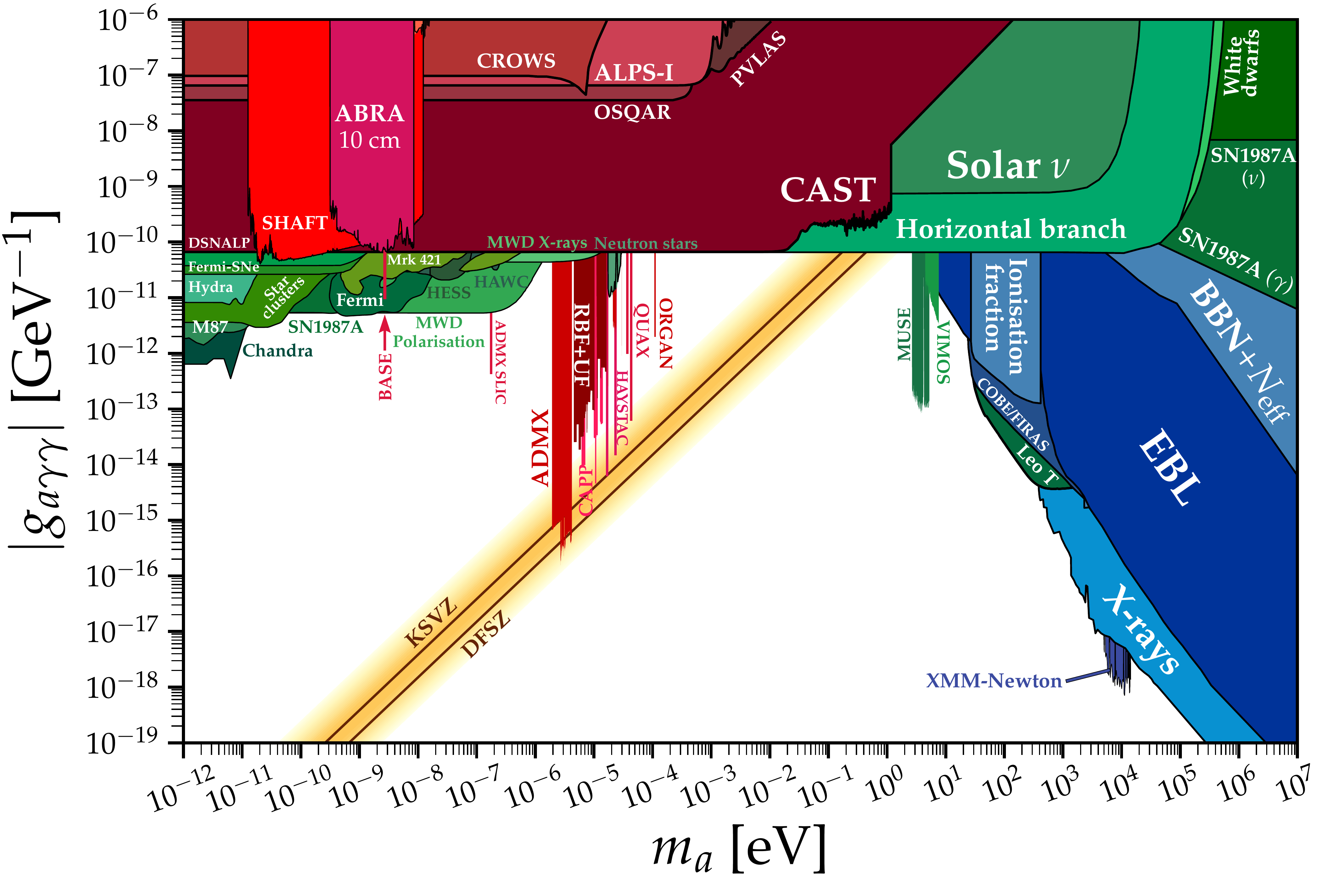}\hspace{2pc}%
		\begin{minipage}[b]{14pc}\caption{\label{fig_ALPs}
				The ALP parameter space, including current experimental bounds and phenomenological constraints (from~\cite{AxionLimits}).}
		\end{minipage}
	\end{figure}
	
	The physics of general ALPs is quite less constrained than the physics of QCD axions, and thus very rich, phenomenologically. 
	Because of space, here we will just overview some characteristics of the ALP parameter space and provide an overview
	of the experimental landscape. 
	
	The ALP landscape, including experimental and other phenomenological bounds, is shown in figure~\ref{fig_ALPs}.
	Several features in this parameter space can be understood from the axion to photon conversion probability in an external magnetic field~\cite{Anselm:1985obz,Raffelt:1987im}
	\begin{align}
		P_{a\gamma}=\left(\frac{g_{a\gamma} B L}{2}\right)^2
		\frac{\sin^2(qL/2)}{(qL/2)^2} \,,
	\end{align}
	with $ q $ the momentum transferred $ q=q_\gamma-q_a \simeq (m_a^2-m_\gamma^2)/2\omega$ (in the last step we took the relativistic limit).
	Let's consider, for example, the case of heliscopes, which search for axions produced in the core of the sun by converting them into photons in a laboratory magnetic field. 
	Solar axions have $ \sim $keV energies.
	Thus, for a magnet of size $L\sim 10\,$m, we expect $ qL\ll 1 $ if $m_a\lesssim 10\,$meV.
	Above this mass, the conversion probability drops as $1/q^2\sim 1/m_a^4$.
	This feature is evident in the sensitivity of CAST~\cite{CAST:2017uph}, currently the most advanced example of axion helioscope, shown in the figure in dark red. 
	Similar arguments apply also to various laboratory experiments such as OSQAR~\cite{Ballou:2015cka} and PVLAS~\cite{DellaValle:2015xxa},
	which at the moment reach a significantly lower sensitivity. 
	
	The same arguments can also help explain 
	the very strong, but limited to small mass, bounds from astrophysical observations, featured in the left side of the parameter space in figure~\ref{fig_ALPs}.
	These bounds are based on the axions oscillation into photons in the galactic magnetic field, and can thus benefit from a very large value of $L$. 
	Notice that the oscillation probability increases quadratically with the magnetic field length (providing B is coherent over such a distance) 
	but that a large $L$ may guarantee coherent conversion only for very small axion masses. 
	
	The central section of the parameter space is dominated by the HB bound.
	In this case, the axion is generated from thermal photons scattering off the electric field of ions and electrons in the core of the star,
	contributing to the stellar energy loss and thus accelerating its evolution. 
	Thus, the weakening of the bound, at $m_a\sim$ a few 10 keV, is not due coherence effects of axion conversion but mostly to thermal (Boltzmann) suppression of the rates for $m_a\gg T_{\rm core}~\sim 10\,$keV~\cite{Cadamuro:2011fd,Carenza:2020zil,Lucente:2022wai}.
	
	Several of the constraints at still larger masses are of astrophysical and cosmological origin, and depend on the axion decay into two photons. 
	The decay probability is proportional to $g_{a\gamma}^2m_a^3$, explaining the rapid strengthening of the bounds at large masses. 
	
	Another notable feature in the ALP parameter space consists in narrow and deep bands from various haloscope experiments, shown in dark red. 
	Haloscopes are experiments searching for axion DM.
	The conventional haloscope technique is the resonant cavity~\cite{Sikivie:1983ip}, 
	which employs the conversion of DM axions in a strong magnetic field that permeates a resonant cavity.
	The conversion is resonant if the axion energy $\omega_a= m_a(1+O(10^{-6}))$ matches a cavity mode.
	These experiments can reach very high sensitivities but at the price of a very 
	slow mass-scanning time, which scales quadratically with the desired signal ($S$) to noise ($N$) level, $\Delta t\propto (S/N)^2$.
	The most mature axion haloscope is the Axion Dark Matter eXperiment (ADMX), which has already reached the standard QCD band in the mass range $m_a\in [2.8 –- 4.2 ]\mu{\rm eV}$ (see Refs.~\cite{ADMX:2018gho,ADMX:2019uok,ADMX:2021nhd}).
	Other cavity experiments are currently probing higher masses. 
	These include HAYSTACK~\cite{Brubaker:2016ktl}, currently probing masses about one order of magnitude larger than ADMX, 
	QUAX--$a\gamma$, probing the mass around $m_a=43~\mu$eV~\cite{Alesini:2020vny},
	RADES, which has provided a limit at $m_a=34.67 \mu$eV~\cite{CAST:2020rlf},
	and ORGAN~\cite{McAllister:2017lkb}, which has a considerably higher mass target $\sim 60-210\mu$eV. 
	
	Considerably different haloscope concepts are currently operating at lower masses.
	In particular, BASE~\cite{Devlin:2021fpq}, SHAFT~\cite{Gramolin:2020ict}, and ABRACADABRA~\cite{Ouellet:2018beu}
	have already reported results in the mass range $m_a\sim 10^{-11}- 10^{-8}\,$eV.

	\section{A look at the Future}
	\label{sec:future}
	
	The experimental searches for axions and ALPs have been expanding very consistently in the last few years, and several new experiments have been proposed and are expected to begin operations in this decade. 
	It is impossible to give a quick overview of the proposed experimental landscape for ALPs, even if we focus solely on the axion-photon coupling. 
	Extremely useful updated plots are publicly available at~\cite{AxionLimits}.
	
	A summary of the present and future searches in the case of QCD axions satisfying relation~\eqref{eq:ma_fa} between mass and PQ constant, is shown in figure~\ref{fig_axion_space_complete}.
	In this figure, exclusion limits are enclosed in continuous lines while expected future sensitivities are shown with dashed contours.
	However, the discussion that follows apply as well to the ALP parameter space, figure~\ref{fig_ALPs}. 
	
	According to the current proposals, there is quite some optimism that a very large portion of the axion parameter space will be explored in the near future.
	The experimental panorama is particularly impressive in the case of future haloscope searches. 
	Several experimental proposals such has 
	MADMAX~\cite{TheMADMAXWorkingGroup:2016hpc,Brun:2019lyf}, and 
	a series of ultra low temperature cavity experiments at IBS/CAPP 
	(CULTASK)~\cite{Semertzidis:2019gkj}
	promise the exploration of the mass region up to a few 0.1 meV.
	A similar range of masses could be probed with tunable axion plasma haloscopes~\cite{Lawson:2019brd}, which employ the axion coupling to plasmons.
	Finally, recent proposals such as TOORAD~\cite{Marsh:2018dlj,Schutte-Engel:2021bqm} and BREAD~\cite{BREAD:2021tpx},
	promise to reach higher masses, all the way into the meV range.
	
	The exploration of the lower mass region has also received considerable attention in recent years.
	The original KLASH proposal~\cite{Alesini:2019nzq} was aiming at the mass range $m_a\in [0.3-1]\mu$eV,
	though some changes in the adopted magnet (from the KLOE to the
	FINUDA magnet) may likely convert it into FLASH, with a slightly higher mass range~\cite{Gatti:2021cel}.
	Furthermore, running experiments such as ABRACADABRA are also expected to reach very high sensitivity, down to the preferred QCD axion band, in the sub-$ \mu $eV mass range~\cite{Kahn:2016aff}.
	
	Meanwhile, the ALPS II laboratory experiment~\cite{Bahre:2013ywa} and, even more so, new axion helioscope proposals BabyIAXO~\cite{IAXO:2020wwp,Dafni:2021mqa} and IAXO~\cite{IAXO:2019mpb} promise to push the search below the astrophysical bounds for a wide range of axion masses.
	If axions do play a role in stellar evolution, there is a good chance that IAXO (and, with some limitations, already BabyIAXO) will be able to see their signatures.

	\section{Discussion and Conclusion}
	\label{sec:conclusion}
	So, will the axion be discovered in the next decade or so? Obviously, it is impossible to know. 
	However, the chances are not so remote. 
	The axion was conceived several decades ago. 
	However, it was realized very soon that phenomenological considerations forced this field to be coupled extremely weakly with SM particles. 
	The term \textit{invisible axion}, adopted in the early times to describe new axion models which could escape the astrophysical bounds, speaks clearly about the expected technical difficulties of realistic axion searches. 
	In spite of groundbreaking progress, especially with the haloscope and helioscope concepts developed by Pierre Sikivie in Ref.~\cite{Sikivie:1983ip}, 
	which presented strategies to \textit{see} the invisible particle,
	it was very clear that finding the axion would require outstanding experimental challenges. 
	It is only now, after several decades of technological development, that the experimental know-how is allowing the exploration of well motivated parameter space. 
	This should be, in my opinion, a reason for enormous excitement and optimism~\cite{Giannotti:2017law}.
	We may not find axions in this decade, but we are definitely going to probe some of the most interesting sections of the axion parameter space.
	
	And, what if we do find the axion?
	Presently, it is difficult to appreciate the full meaning of such a discovery. 
	Axions are excellent CDM candidates.
	However, discovering it will not guarantee that axions are the totality (or even a large fraction) of the CDM in the universe. 
	Furthermore, depending on how the axion will be discovered, it is possible that some important parameters, for example its mass, will be not be immediately discernible. 
	So, an axion signal would likely motivate an enormous experimental effort in determining its properties. 
	This is, obviously, and extremely exciting prospect. 
	
	Even after an hypothetical discovery, direct experimental searches would not be the only way to study axions or ALPs.
	Astrophysics will likely continue to play a fundamental role, 
	especially if axions happen to have relatively large couplings.
	Axions with couplings close to the current bounds might play a very significant role in understanding important properties of the sun, including its metallicity~\cite{Jaeckel:2019xpa}, and its core magnetic field~\cite{Guarini:2020hps,Caputo:2020quz,OHare:2020wum}.
	Very light (sub-neV) axions might provide by far the most efficient messengers to study stars in late evolutionary stages,\footnote{The mass requirement derives from the fact that such stars are too far for a direct detection of the axions. For a detectable signal, axions would have to convert into photons in the galactic magnetic field.}
	possibly providing insights on the particular evolutionary phase, on nuclear core reactions, and on the time till core collapse~\cite{Xiao:2020pra,Xiao:2022rxk}.
	
	After decades of research and technological developments, the times are mature for a fundamental breakthrough in particle physics, astrophysics and cosmology. 
	Perhaps, it will be an axion discovery.

	\ack 
	The author wishes to thank the Large TPC conference organizers for the opportunity to participate and present this overview.
	I am indebted to C. O'Hare for providing me (and the rest of the world) with updated and carefully presented maps of the ALP parameter space. 
	Warm thanks to my several collaborators, in particular Luca Di Luzio, Igor G. Irastorza, Giuseppe Lucente, Federico Mescia, and Alessandro Mirizzi who read and provided comments on this draft. 
	
	\vspace{1cm}

\bibliographystyle{JHEP.bst}
\bibliography{paper_Arxiv.bib}

\end{document}